\documentstyle[a4,12pt,epsf,amssymb]{article}
\begin{document}
\textheight 23cm
\topmargin -0.5cm
\textwidth 19cm
\oddsidemargin 0.65cm
\evensidemargin 0.65cm
\setlength{\parskip}{0.45cm}
\setlength{\baselineskip}{0.75cm}
%
\begin{titlepage}
\begin{flushright}
TPR-98-22\\
December 1998\\
(revised version)
\end{flushright}
\begin{center}
\renewcommand{\thefootnote}{\fnsymbol{footnote}}
\setcounter{footnote}{1}
\vspace*{-0.1cm}
{\LARGE
{\bf Numerical Solution of the Evolution}  

\vspace{-0.2cm}
{\bf Equation for Orbital Angular} 

\vspace{-0.2cm}
{\bf Momentum of Partons in the Nucleon  }
}
\vspace{0.1cm}

\vspace*{0.1cm}
{\Large O.~Martin, P. H\"agler, A.~Sch\"afer}
\vspace{0.1cm}
\linebreak
Institut f\"ur Theoretische Physik, Universit\"at Regensburg, \\
D-93040~Regensburg, Germany
\vspace*{0.5cm}
\linebreak
\begin{flushleft}
\end{flushleft}
\renewcommand{\thefootnote}{\fnsymbol{roman}}
\setcounter{footnote}{0}
{\bf Abstract}

\vspace{-0.3cm}
\end{center}
The evolution of orbital angular momentum distributions
within the radiative parton model is studied. We use different 
scenarios for the helicity weighted parton distributions and
consider a broad range of input distributions for
orbital angular momentum. In all cases we are lead to the
conclusion that the absolute value of the average orbital angular momentum
per parton peaks at relatively large $x\approx 0.1$
for perturbatively accessible scales. Furthermore, 
in all scenarios considered here the average orbital momentum 
per parton is several times larger for gluons than for quarks which
favours gluon initiated reactions to measure orbital angular
momentum. The large gluon polarization typically obtained in NLO-fits to DIS data
is in each case primarily canceled by the gluon orbital angular momentum.
\end{titlepage}

\section{Introduction}
The correct treatment of orbital angular momentum is one of the 
theoretically most important and probably also most controversal 
problems in hadron spin physics [1-13]. On the one hand the ususal NLO-fits
to
deep inelastic data predict a very large gluon spin (of order 1
$\hbar$)
which has to be balanced by a correspondingly large negative orbital
angular momentum contribution. As this orbital angular momentum results
in substantial transverse momentum with respect to the spin
orientation, it should be e.g. observable in specific 
semi-inclusive reactions with transversely polarized nucleons.
Thus it is not just a formal construction of the theoretical
description, but a real physical quantity. On the other hand
the most natural definition of angular momentum is not gauge
invariant.
Naturally one is free to chose any specific gauge and we will 
actually work in the lightcone gauge $A^+(x)=0$, but even after
fixing this gauge there remains some residual gauge-freedom.
Different attempts to come to terms with this problem 
differ in their interpretation of what is called 
gluon spin, quark orbital angular momentum and gluon orbital angular
momentum. The only uncontroversal quantity is the quark spin
distribution. The case for our choice  (which was first proposed by
Ji et al. \cite{Ji1} and Jaffe et al. \cite{Ja1})
\begin{eqnarray}
 \Delta\Sigma &=& \langle P',S|\int d^3x\,
  \frac{i}{2} \overline{\Psi}\gamma^+[\gamma^1,
  \gamma^2]_-\Psi |P,S\rangle 
 \label{deltasigma} \\ 
  \Delta g &=& \langle P',S|\int d^3x\, (A^1
  \partial^+A^2 - A^2\partial^+A^1)|P,S\rangle 
 \label{deltag} \\
 L_q &=& \langle P',S|\int d^3x\, i \overline{\Psi}
  \gamma^+ (x^1\partial^2-x^2\partial^1)\Psi |P,S
  \rangle 
 \label{Lq} \\
  L_g &=& \langle P',S| \int d^3x\, \partial^+
 A^j (x^1\partial^2-x^2\partial^1) A^j
 |P,S \rangle
 \label{Lg}
\end{eqnarray} 
was recently
substantially
strengthened by Bashinsky and Jaffe \cite{Ja5}. 
They proposed a gauge-invariant 
formulation which in the $A^+(x)=0$ gauge reduces to the form we used.
In doing so they could specify that the residual gauge-freedom 
can be fixed by the additional constraint $A^{\mu}(x_{\nu}\rightarrow
\infty)=0$. Based on their work we can therefore conclude that 
our treatment gives results which can be interpreted in a
straightforward manner (i.e. our orbital angular momentum corresponds
really to the naive interpretation) up to effects related to gauge
field configurations for which the  combined gauge
condition 
\begin{equation}
A^+(x)=0~~~~~~~~~~~{\rm and} ~~~~~~~~~~~A^{\mu}(x_{\nu}\rightarrow
\infty)=0
\end{equation}
does not apply. (We are not aware of any physically sensible 
gauge-field configuration for which the choice (1) would not be
possible.)\\

In \cite{haegl} we derived the complete coupled evolution equations 
for all moments, which read:
\begin{eqnarray}
\label{evoequ}
&&\frac{d}{dt}
\left(
\begin{array}{llll}
 \Delta\Sigma^n(t)\\
 \Delta g^n(t) \\
 L_q^n(t) \\
 L_g^n(t) \\
\end{array}
\right)
= \frac{\alpha(t)}{2\pi} 
\left(
\begin{array}{cc}
 A^n_{SS} & A^n_{SL} \\
 A^n_{LS} & A^n_{LL}
\end{array}
\right)
\left(
\begin{array}{llll}
 \Delta\Sigma^n(t) \\
 \Delta g^n(t) \\
 L_q^n(t) \\
 L_g^n(t) \\
\end{array}
\right)
\\ 
\nonumber\\
&&A^n_{SS}=\left(
\begin{array}{cc}
 C_F\left[\frac{3}{2}+\frac{1}{n(n+1)}-2\sum^n_{j
 =1}\frac{1}{j}\right] & n_f\left[\frac{n-1}{n(n
 +1)}\right] \\
 C_F\left[\frac{n+2}{n(n+1)}
\right] & 2C_A\left[\frac{11}{12}-
\frac{n_f}{6C_A}+\frac{2}{n(n+1)}-\sum^n_{j=1}\frac{1}{j}\right]
\end{array}
\right)
\nonumber \\
&&A^n_{SL}=\left(
\begin{array}{cc}
 0 & 0 \\
 0 & 0
\end{array}
\right)
\nonumber \\
&&A^n_{LS}=\left(
\begin{array}{cc}
 -2C_F\left[\frac{1}{n(n+2)}\right] & n_f\left[
\frac{n^2+n+6}{n(n+1)(n+2)(n+3)}\right] \\
 -C_F
\left[\frac{n+4}{n(n+1)(n+2)}\right] & -4C_A
\left[\frac{n^2+4n+6}{n(n+1)(n+2)(n+3)}
\right]
\end{array}
\right)
\nonumber \\
&&A^n_{LL}=\left(
\begin{array}{cc}
 C_F\left[\frac{3}{2}-\frac{2n
  +3}{(n+1)(n+2)}-2
\sum^n_{j=1}\frac{1}{j}\right] & n_f\left[
\frac{n^2+3n+4}{(n+1)(n+2)(n+3)}\right] \\
 C_F
\left[\frac{n^2+3n+4}{n(n+1)(n+2)}\right] & 2C_A
\left[-\frac{n^3+3n^2-6}{n(n+1)(n+2)(n+3)}-
\frac{n_f}{6C_A}+\frac{11}{12}-\sum^n_{j=1}
\frac{1}{j}\right]
\end{array}
\right)
\nonumber
\end{eqnarray}
In this contribution we present results of numerical studies for
these evolution equations which were obtained using the Mellin method
\cite{GRV90}.

\section{Results}
A study of the evolution of orbital angular momentum requires the knowledge
of the helicity weighted parton distributions, which have not yet been
determined with sufficient precision. 
We take this uncertainty into account by using two different sets
of polarized distributions, namely the GRSV leading-order standard and gmax
scenario
\cite{glre3}. For both sets analytic expressions of the polarized parton
distributions
are given at a very low hadronic scale of $\mu_0^2=0.23$~GeV$^2$. The GRSV
standard
scenario has the property that $\frac{\Delta \Sigma(\mu_0^2)}{2}+\Delta
g(\mu_0^2)\approx
0.475$\footnote{Throughout the paper we adopt the shorter notation 
$\Delta f(\mu^2)$ for the first moment $\Delta f^1(\mu^2)$ of the distribution
$\Delta f(x,\mu^2)$.} so that according to the spin sumrule 
\begin{equation}
\frac{\Delta \Sigma(\mu^2)}{2}+\Delta g(\mu^2) +L_q(\mu^2)+ L_g(\mu^2)
= \frac{1}{2}
\end{equation}
quarks and gluons carry barely any orbital angular momentum at the 
initial scale. This means that
predictions for this scenario should be largely independent on the
choice of $L_g(x,\mu_0^2)$ and $L_q(x,\mu_0^2)$. While in the
standard scenario $\Delta g(\mu^2)$ is moderately positive, the main features
of the gmax scenario are a saturated polarized gluon distribution 
$\Delta g(x,\mu_0^2)=g(x,\mu_0^2)$ and large initial orbital angular momentum
$L_q(\mu_0^2)+L_g(\mu_0^2)=-0.43$. Therefore, the gmax scenario is 
ideal to study the dependence of the evolution results on 
the shape and size of $L_g(x,\mu_0^2)$ and $L_q(x,\mu_0^2)$.

\begin{figure}
\begin{center}
\epsfysize19cm
\leavevmode\epsffile{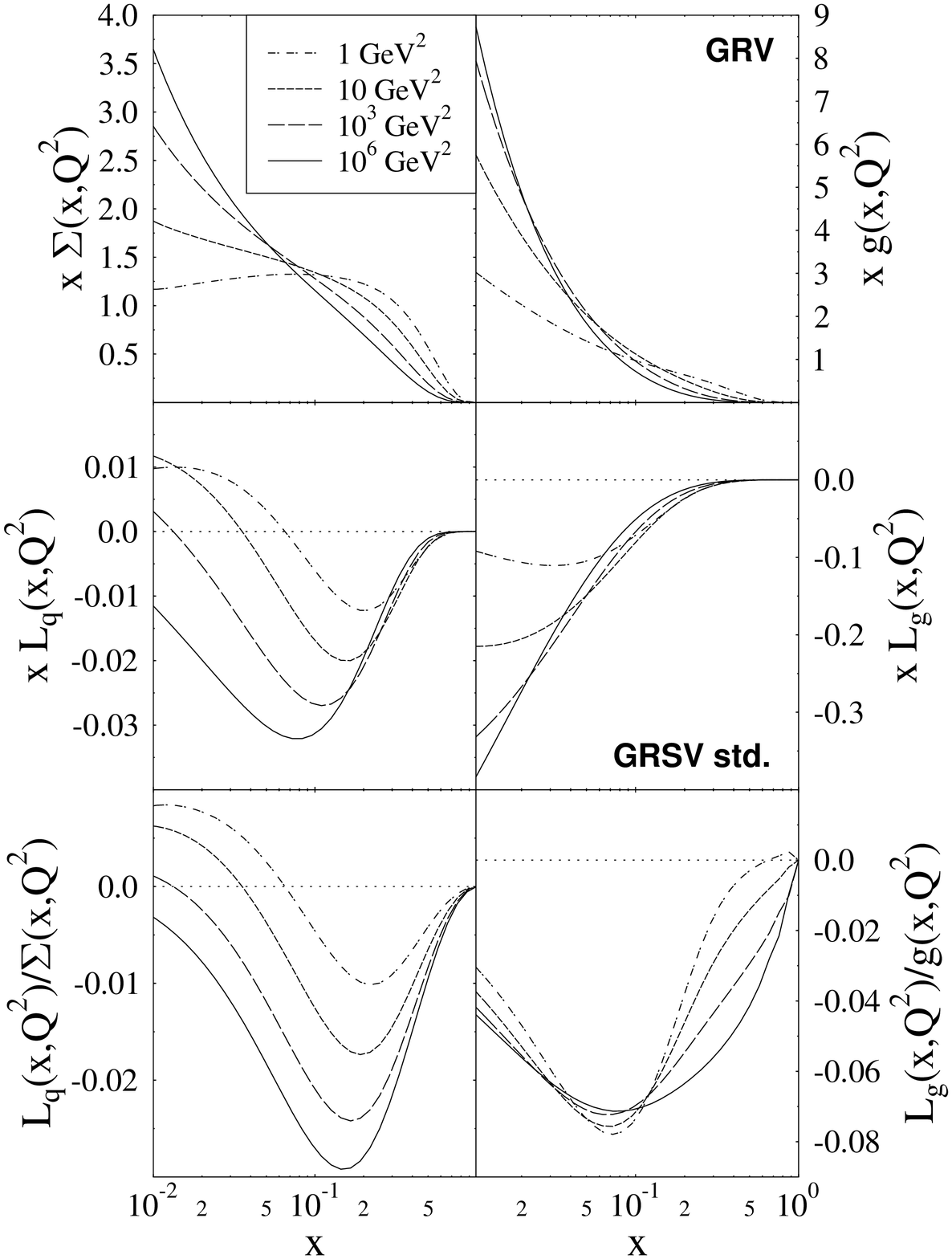}
\caption{\label{fig1}
Evolution of orbital angular momentum. The GRSV LO standard scenario is
used as input for the polarized parton distributions at $\mu_0^2=0.23$~GeV$^2$.
At this scale the missing angular momentum of ~0.025 units is evenly
distributed among $L_q(x,\mu_0^2)$ and $L_g(x,\mu^2)$, which are assumed to
have the same shape as $u_v(x,\mu_0^2)$ and $g(x,\mu_0^2)$, respectively.
}
\end{center}
\end{figure}

\begin{figure}
\begin{center}
\epsfysize13cm
\leavevmode\epsffile{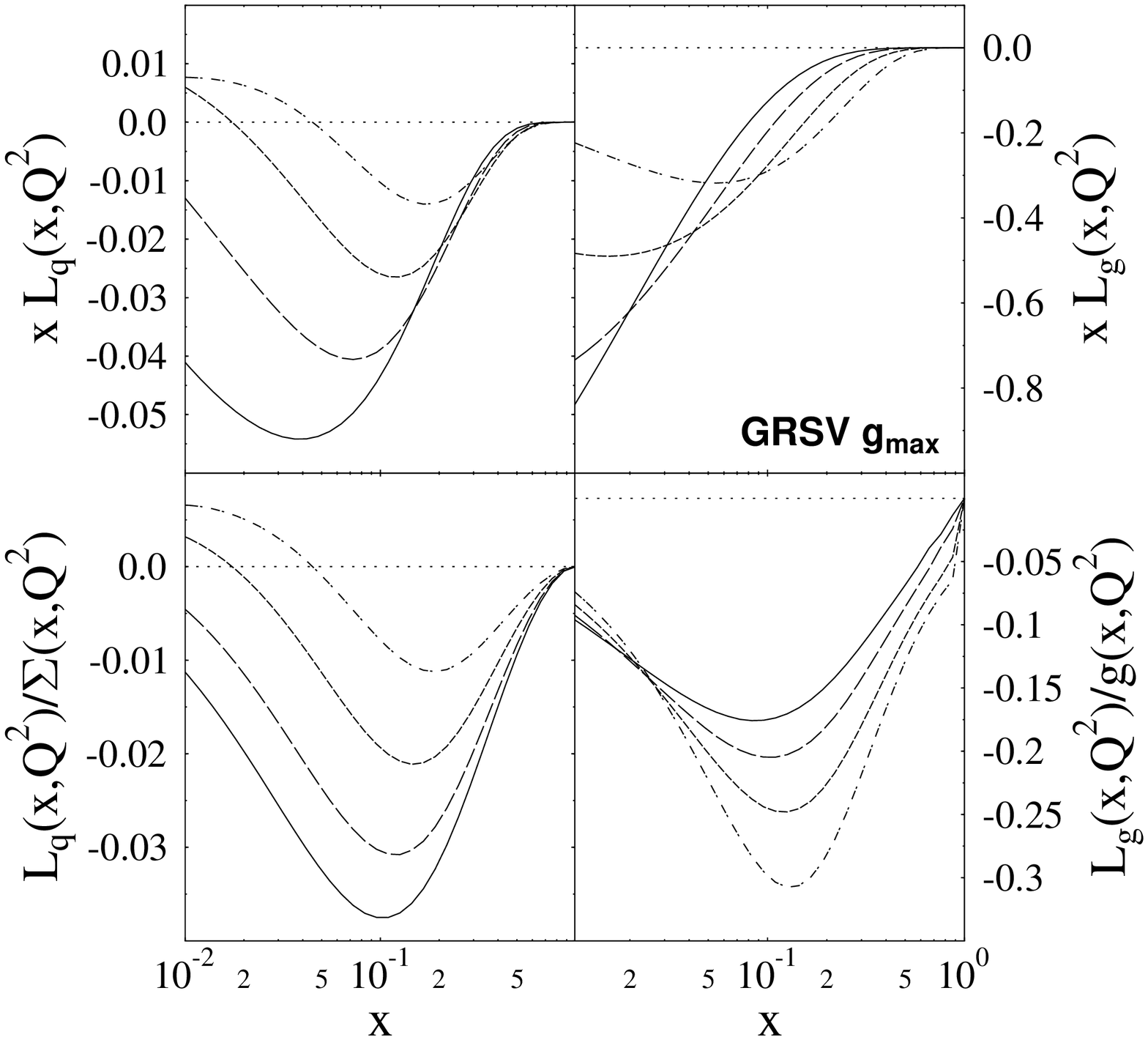}
\caption{\label{fig2}
Evolution of orbital angular momentum. The GRSV LO gmax scenario is
used as input for the polarized parton distributions at $\mu_0^2=0.23$~GeV$^2$.
The missing angular momentum of -0.43 units is evenly
distributed among $L_q(x,\mu_0^2)$ and $L_g(x,\mu^2)$, which again are assumed
to
be proportional to $u_v(x,\mu_0^2)$ and $g(x,\mu_0^2)$, respectively.
}
\end{center}
\end{figure}

Fig. \ref{fig1} and \ref{fig2} show the quark and gluon orbital angular 
momentum evolved to various scales ranging from 1~GeV$^2$ to $10^6$~GeV$^2$
within the standard and gmax scenario, respectively.
It is reasonable to assume that at a low hadronic scale the quark
orbital 
angular
momentum is mainly carried by valence quarks so that $L_q(x,\mu_0^2)$ has been
chosen to be proportional to $u_v(x,\mu_0^2)$ and also $L_g(x,\mu_0^2)\propto
g(x,\mu_0^2)$. Additionally, we distributed the initial orbital angular momentum
evenly between quarks and gluons such that $L_q(\mu_0^2)=L_g(\mu_0^2)$.
The figures show that $L_q(\mu^2)$ and $L_g(\mu^2)$ are negative and
decrease for growing $\mu^2$. 
This behaviour has been expected since the
quark axial charge is conserved under leading-order evolution whereas 
$\Delta g(\mu^2)$ is positive and grows approximately like
$\alpha_s^{-1}(\mu^2)$. Therefore, 
the total orbital angular momentum must decrease when $\mu^2$ increases.
Fig.~\ref{fig1}
and \ref{fig2} also show that the average orbital angular momentum per parton
($L_q(x,\mu^2)/\Sigma(x,\mu^2)$ and $L_g(x,\mu^2)/g(x,\mu^2)$) has 
its maximum at a relatively large $x$-value of approximately 0.1. Additionally,
in both scenarios gluons carry far more orbital angular momentum
per parton than quarks do. 

\begin{figure}
\begin{center}
\epsfysize14cm
\leavevmode\epsffile{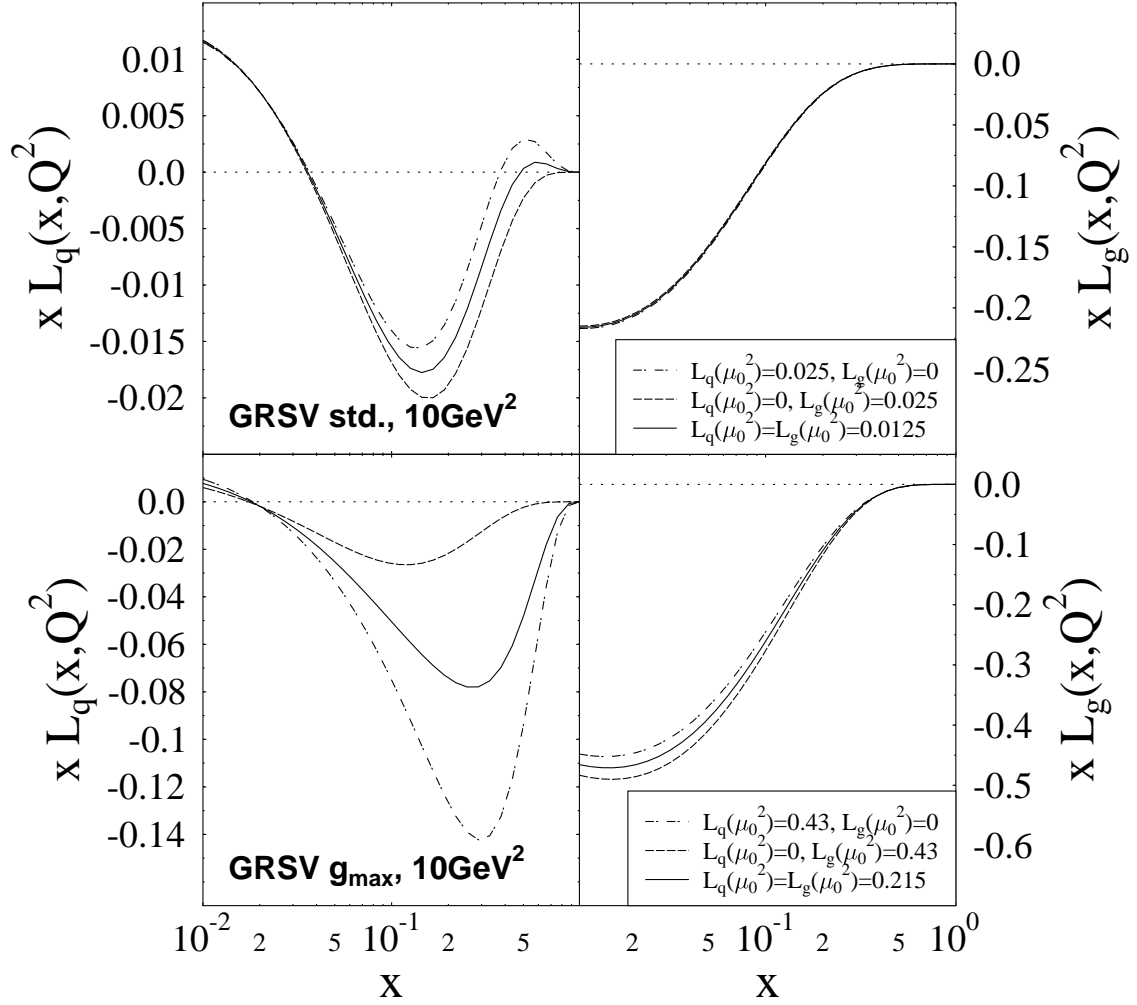}
\caption{\label{fig3}
The missing angular momentum is distributed among $L_q(x,\mu_0^2)$
and $L_g(x,\mu_0^2)$ in three different ways, by setting the first
moments of these distributions either equal or to 0. 
}
\end{center}
\end{figure}

\begin{figure}
\begin{center}
\epsfysize14cm
\leavevmode\epsffile{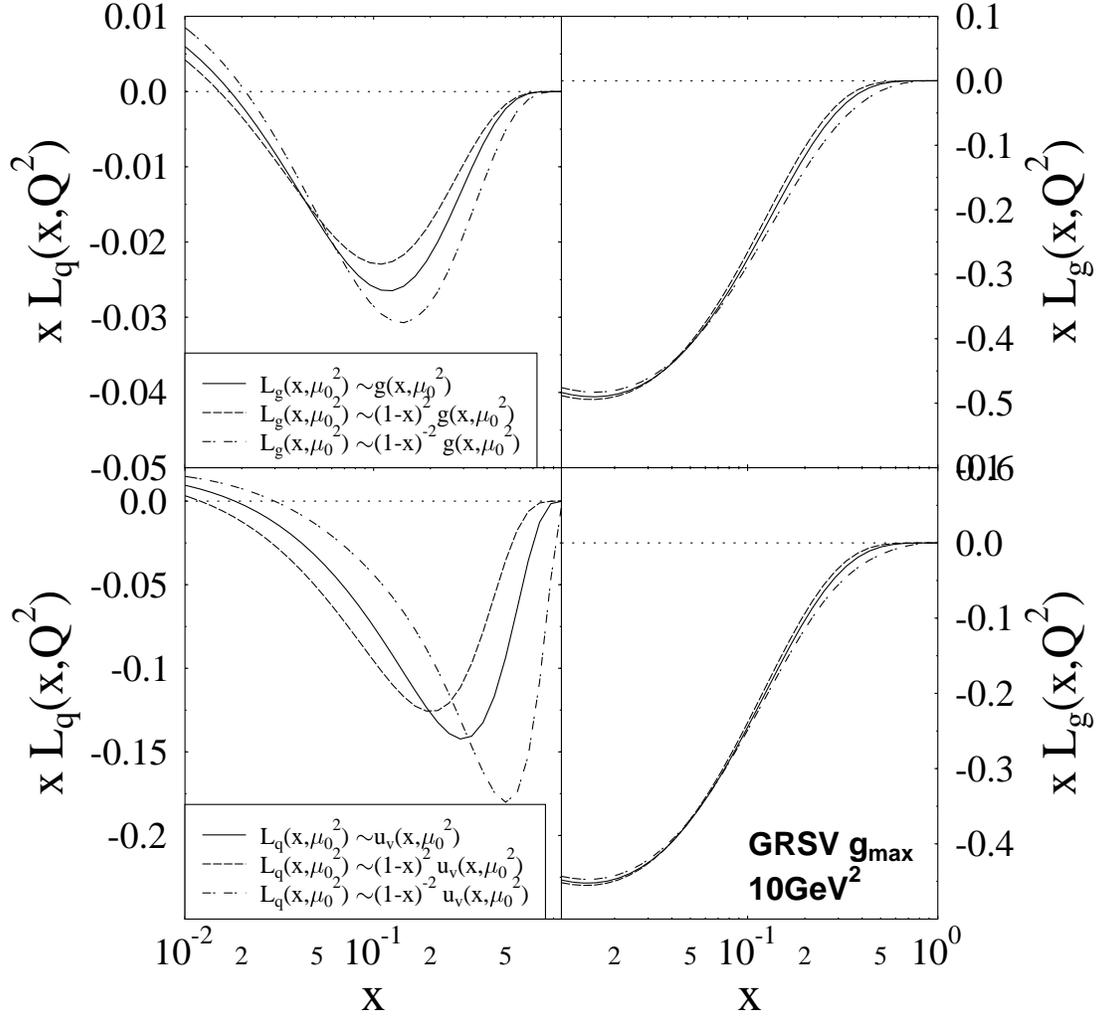}
\caption{\label{fig4}
Dependence of the orbital angular momentum distributions on the shape
of $L_g(x,\mu_0^2)$ and $L_q(x,\mu_0^2)$ at large $x$. 
$L_q(x,\mu_0^2)$ ($L_q(x,\mu_0^2)$) is set to 0 when  
$L_g(x,\mu_0^2)$ ($L_q(x,\mu_0^2)$) is varied.}
\end{center}
\end{figure}

The assumptions made about the initial quark and gluon orbital angular momentum
distributions in fig.~\ref{fig1} and \ref{fig2} are somewhat arbitrary. We
checked therefore  how they affect the evolved distributions at perturbative
scales.
For all following
results we chose the scale to be $\mu^2=10$~GeV$^2$.
In fig.~\ref{fig3} we varied the magnitude of $L_q(x,\mu_0^2)$ and
$L_g(x,\mu_0^2)$
by setting one of the distributions to 0 and attributing all of the missing
angular momentum to the other. While this
does not lead to a significant change of
$L_g(x,\mu^2)$ in both scenarios, $L_q(x,
\mu^2)$ varies approximately by a factor of
7 in the gmax scenario. We also checked the dependence of the evolved orbital
angular distributions on the shape of $L_q(x,\mu_0^2)$ and $L_g(x,\mu_0^2)$ at 
large $x$. Only results for the gmax scenario are shown because according to
fig.~\ref{fig3} the results in the standard scenario depend
only weakly on the initial orbital angular momentum distributions.
In the upper part of fig.~\ref{fig4} we set $L_q(x,\mu_0^2)=0$ and took
$L_g(x,\mu_0^2)$ to be proportional to $g(x,\mu_0^2)$, $(1-x)^2g(x,\mu_0^2)$
and 
$(1-x)^{-2}g(x,\mu_0^2)$, in the lower part we similarly varied the shape
of $L_q(x,\mu_0^2)$ with $L_g(x,\mu_0^2)=0$. 
Again, only the quark orbital angular momentum distribution shows a significant
dependence on the initial shape which is stronger when $L_q(x,\mu_0^2)$ is
changed.

\begin{figure}
\begin{center}
\epsfysize8cm
\leavevmode\epsffile{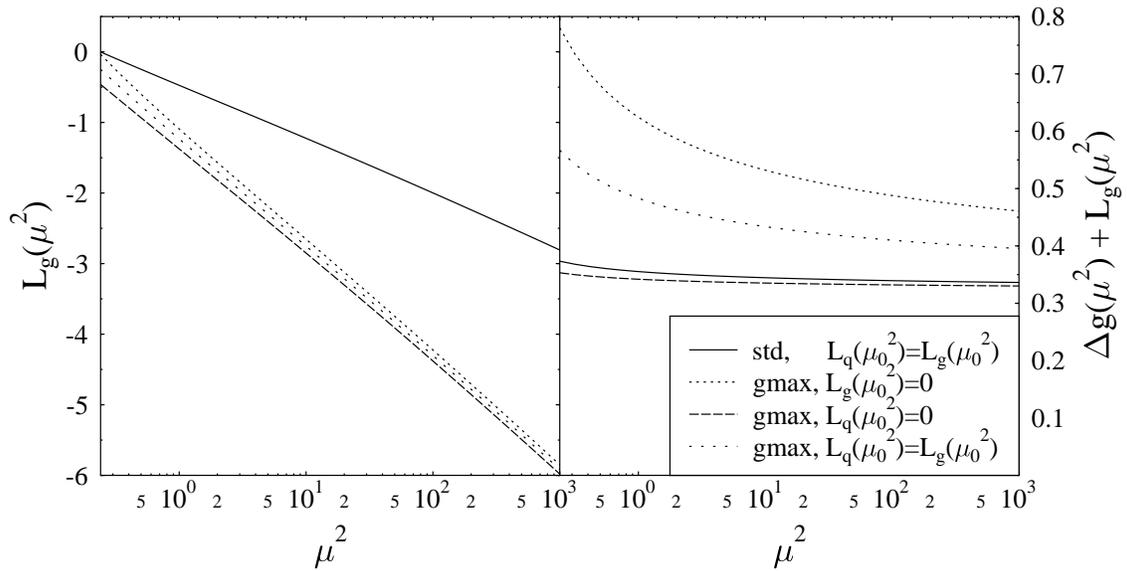}
\caption{\label{fig5}
Evolution of the first moment of $L_g(x,\mu^2)$ and the gluon contribution
to the proton spin in different scenarios. }
\end{center}
\end{figure}

Fig.~\ref{fig5} gives a clue on why the gluon orbital angular distribution
is so large and on why it basically only depends on the polarized parton
distributions.
We see that, even though the orbital angular momentum carried by gluons becomes
rapidly large when the scale increases, the total contribution of the gluons to
the
spin of the proton remains relatively stable. This means that 
$L_g(\mu^2)$ behaves similar to  $-\Delta g(\mu^2)$.  
Indeed this behaviour is reflected by the anomalous dimension matrix for
the first moments. It has two unusually large entries, namely
$(A^1_{SS})_{22}\approx 4.5$, 
which leads to the built-up of a large and positive $\Delta g(\mu^2)$, and
$(A_{LS})_{22}=-4.5$, 
which leads to the observed coupling of the gluon orbital angular momentum
to the polarized gluon distribution. In order to test our evolution code we also
calculated the analytical solution of the evolution equation for the first 
moments of the orbital
angular momentum distributions:
\begin{eqnarray}
L_q(t)&=&\left(\frac{t}{t_0}\right)^{-\frac{2(16+n_f)}{3(33-2n_f)}}
  \left(L_q(t_0)+\frac{\Delta\Sigma(t_0)}{2}
-\frac{3n_f}{2(16+3n_f)}\right)
\nonumber\\&&
  -\frac{\Delta\Sigma(t_0)}{2}+\frac{3n_f}{2(16+3n_f)},
\\
L_g(t)&=&\left(\frac{t}{t_0}\right)^{-\frac{2(16+n_f)}{3(33-2n_f)}}
  \left(L_g(t_0)+\Delta g(t_0)-\frac{16}{2(16+3n_f)}\right)
\nonumber\\&&
  -\Delta g(t)+\frac{16}{2(16+3n_f)}. \label{lgt}
\end{eqnarray}
Both quantities have a contribution which vanishes like a negative power of $t$ for
$\mu^2\rightarrow \infty$ and a also a constant term. However, the $-\Delta g(t)$
contribution only appears in equation~(\ref{lgt}) which is consistent with
our numerical results.

\section{Conclusions}
We studied the evolution of orbital angular momentum in the
GRSV standard and gmax scenario for a variety of
input distributions $L_q(x,\mu_0^2)$ and $L_g(x,\mu_0^2)$.
The ratio of both distribution functions to the unpolarized parton 
distribution functions peak  at relatively large 
$x\approx 0.1$ at perturbatively
accessible scales. This result sustains the hope to find signs of
orbital angular momentum  for example in semi-inclusive reactions.
Gluon initiated reactions might be better suited since the average orbital
angular momentum per parton is several times larger for gluons than for quarks
in all
scenarios considered here. (However, one should keep in mind that 
$L_g(\mu^2)$ is closely coupled to $\Delta g(\mu^2)$ by the evolution
equations so that its dominance over $L_q(\mu^2)$ should be less 
pronounced for very small $\Delta g(\mu^2_0)$.)
We found furthermore that $L_g(\mu^2)$ and $\Delta g(\mu^2)$ cancel to
a large extent.
$L_g(x,\mu^2)$ is 
much more stable  under variation of the input
distributions for orbital angular momentum than $L_q(x,\mu^2)$. 
It almost exclusively depends on the
polarized quark singlet and gluon distributions. Thus the
radiative parton model should successfully predict $L_g(x,\mu^2)$
once the polarized distribution functions are determind with good precision.

\section*{Acknowledgements}
We
acknowledge financial support from the BMBF and the Deutsche 
Forschungsgemeinschaft and thank Bob Jaffe for very helpful
discussions.
We are also  indebted
to T.~Gehrmann for providing us with an evolution program for
polarized parton distributions.



\end{document}